\documentclass[amsmath,amssymb]{revtex4}
\usepackage{graphicx}% Include figure files
\usepackage{dcolumn}% Align table columns on decimal point
\usepackage{bm}% bold math

\newtheorem{thm}{Theorem}[subsection]
\usepackage{amsfonts}
\usepackage{graphicx, color}
\usepackage{hyperref}

\begin{document}

% Use the \preprint command to place your local institutional report
% number in the upper righthand corner of the title page in preprint mode.
% Multiple \preprint commands are allowed.
% Use the 'preprintnumbers' class option to override journal defaults
% to display numbers if necessary
%\preprint{}

%Title of paper
\title{A mathematical proof for a ground-state identification criterion}

% repeat the \author .. \affiliation  etc. as needed
% \email, \thanks, \homepage, \altaffiliation all apply to the current
% author. Explanatory text should go in the []'s, actual e-mail
% address or url should go in the {}'s for \email and \homepage.
% Please use the appropriate macro foreach each type of information

% \affiliation command applies to all authors since the last
% \affiliation command. The \affiliation command should follow the
% other information
% \affiliation can be followed by \email, \homepage, \thanks as well.
\author{Tien D. Kieu}
\email{kieu@swin.edu.au}
%\homepage[]{Your web page}
% \thanks{}
%\altaffiliation{}
\affiliation{Centre for Atom Optics and Ultrafast Spectroscopy, ARC
Centre of Excellence for Quantum-Atom Optics, Swinburne University
of Technology, Hawthorn 3122, Australia}

%Collaboration name if desired (requires use of superscriptaddress
%option in \documentclass). \noaffiliation is required (may also be
%used with the \author command).
%\collaboration can be followed by \email, \homepage, \thanks as well.
%\collaboration{}
%\noaffiliation

%\begin{abstract}
%\end{abstract}

% body of paper here - Use proper section commands
% References should be done using the \cite, \ref, and \label commands
%\section{\label{}}
% Put \label in argument of \section for cross-referencing
%\section{\label{}}
%\subsection{}
%\subsubsection{}
% \date{\today}

% \section*{Introduction and outline of the paper}
%%%% Review of other bounds??
%One of the most important problems of contemporary mathematics and
%computer science is whether $NP\stackrel{?}{=}P$~\cite{Clay}. As
%classical computation has so far failed to come up with a tractable
%method to deal with NP problems, some hope has been pinned on the
%alternative quantum computation.  And yet, despite the discoveries
%of important algorithms for factoring~\cite{Shor} and unstructured
%searching~\cite{Grover}, quantum computation, in its most general
%form, has not been successful either in the delivery of any improved
%and superior algorithm.
\begin{abstract}
We give a mathematical proof for an identification criterion by a
probability measure for the ground state among an infinite number of
available states, or a finitely truncated number with appropriate
boundary conditions, in a quantum adiabatic algorithm for Hilbert's
tenth problem.
\end{abstract}

\date{February 19, 2006}

\maketitle

\section*{Background}
In a quantum adiabatic algorithm for Hilbert's tenth
problem~\cite{kieuFull}, we have provided a mathematical proof in
two dimensions for an identification criterion for the ground state
of the final Hamiltonian.  The criterion states that, in the case of
no degeneracy for the final Hamiltonian $H_P$, the Fock state that
has a measurement probability of greater than one-half {\em is} the
ground state of $H_P$.	At first, we thought that the proof was also
valid for higher dimensions.	Thanks to Warren Smith, who provided a
counterexample in five dimensions~\cite{Smith2005}, we now know that
our proof is insufficient for higher dimensions. However, we have
previously pointed out in our reply (in an earlier version
of~\cite{kieuNew}): (i) that such a counterexample by Smith could be
realised only because of the artefact resulted from the truncation,
and its associated boundary conditions, in the dimensions of a
dimensionally infinite Hilbert space; (ii) that a suitable choice of
some complex parameters ($\alpha$'s in~(\ref{H_I}) below) would
restore the identification criterion even for finite dimensions;
(iii) and that we suspected, nevertheless, that our criterion would
still be valid in {\em infinite} dimensions.  In this short note, we
present a mathematical proof for this last claim, and also for the
case of finitely truncated Hilbert spaces with appropriate boundary
conditions.

\section*{Introduction}
Our quantum adiabatic algorithm~\cite{kieuFull} for a Diophantine
equation, with $K$ variables,
\begin{eqnarray}
D(x_1, \ldots, x_K) &\stackrel{?}{=} 0, \label{Diophantine}
\end{eqnarray}
initially starts with a coherent state,
\begin{eqnarray}
|\psi(0)\rangle &=& \bigotimes^K_{i=1}\left\{{\rm
e}^{-|\alpha_i|^2/2}\sum_{n_i=0}^\infty
\frac{\alpha_i^{n_i}}{\sqrt{n_i!}} |n_i\rangle\right\},
\label{coherentstate}
\end{eqnarray}
which is the nondegenerate ground state of a Hamiltonian $H_I$, with
complex numbers $\alpha_i\not= 0$,
\begin{eqnarray}
H_I &=& \sum_{i=1}^K \left(a^\dagger_i - \alpha_i^*\right)\left(a_i
- \alpha_i\right). \label{H_I}
\end{eqnarray}
This Hamiltonian is then linearly extrapolated in a time $T$ via a
time-dependent Hamiltonian ${\cal H}(t)$,
\begin{eqnarray}
{\cal H}(t) &=& H_I + \frac{t}{T}(H_P-H_I), \label{calH}
\end{eqnarray}
to a final Hamiltonian $H_P$,
\begin{eqnarray}
H_P &=& \left(D(a^\dagger_1 a_1, \ldots, a^\dagger_K a_K)\right)^2,
\label{H_P}
\end{eqnarray}
which encodes in its ground state the information about the
existence of solution of the Diophantine
equation~(\ref{Diophantine})~\footnote{For simplicity, we assume
that the ground state here is nondegenerate; see~\cite{kieuFull} for
the general situation.}.

In order to have the measurement probability of {\em any} excited
state of $H_P$ less than one-half for {\em any} time interval $T$,
we can derive from~\cite{kieuFull} as part of the sufficient
conditions the requirement that
\begin{eqnarray}
\langle e(t)| H_P - H_I |f(t)\rangle &\not=&0,\label{condition}
\end{eqnarray}
for all $0<t<T$, for {\em any} pairs of orthogonal instantaneous
eigenstates $|e(t)\rangle$ and $|f(t)\rangle$ of ${\cal H}(t)$.

Note that from the orthonormal instantaneous eigenstates,
\begin{eqnarray}
&&\|\;|f(t)\rangle\| = \|\;|e(t)\rangle\| = 1, \label{norm}\\
&&\langle e(t)|f(t)\rangle = 0, \label{ortho}\\
&&{\cal H}(t) |e(t)\rangle  = E_e(t) |e(t)\rangle, \;\;{\rm and}
\;\; {\cal H}(t) |f(t)\rangle	 = E_f(t) |f(t)\rangle, \label{eigen1}
\end{eqnarray}
and from the explicit expression for ${\cal H}(t)$ in~(\ref{calH}),
we can easily see that a violation of the
condition~(\ref{condition}) implies
\begin{eqnarray}
\langle e(t_0)| H_P - H_I |f(t_0)\rangle = 0 &\Leftrightarrow& {\rm
both}\;\;
\langle e(t_0)| H_P |f(t_0)\rangle =0, \label{cond1}\\
&& {\rm and}\;\; \langle e(t_0)| H_I |f(t_0)\rangle =0,
\label{cond2}
\end{eqnarray}
at some time $t_0$ for at least one pair of instantaneous
eigenstates.

For two dimensions, there are only two eigenstates at each $t$, and
the last two conditions above are unsatisfiable if the two end-point
Hamiltonians do not commute~\cite{kieuFull},
\begin{eqnarray}
[H_I, H_P]&\not=&0.
\end{eqnarray}
For higher number of dimensions, this noncommutativity is no longer
sufficient and we will have to resort to the explicit forms of $H_I$
and $H_P$.

\section*{A mathematical proof for infinite dimensions}
We present here a proof by contradiction that the
condition~(\ref{condition}) always holds with infinite dimensions.
That is, we assume that the conditions~(\ref{cond1})
and~(\ref{cond2}) are satisfied at some time $t_0$, then derive from
them some contradictions in order to conclude that~(\ref{condition})
must hold.	We restrict for simplicity to the case of one variable,
$K=1$ in~(\ref{Diophantine}); and the proof should be generalisable
to other finite values of $K$.

We first expand the instantaneous eigenstates at this time $t_0$ in
terms of the Fock states
\begin{eqnarray}
|f(t_0)\rangle = \sum_{n=0}^\infty f_n |n\rangle, \;\;{\rm and}\;\;
|e(t_0)\rangle = \sum_{n=0}^\infty e_n |n\rangle.
\end{eqnarray}
The conditions~(\ref{ortho}), and~(\ref{cond1}) can then be
represented, respectively, as
\begin{eqnarray}
\sum_{n=0}^\infty e^*_n f_n &=& 0,\label{condA}\\
\sum_{n=0}^\infty \left(D(n)\right)^2 e^*_n f_n &=& 0.\label{condB}
\end{eqnarray}

We next make use of an important theorem~\footnote{Theorem ({\bf
4.2}, I), pp. 65-66, in~\cite{Cooke1950}.}, reproduced here word by
word,
\begin{thm}[Bosanquet-Henstock]
The necessary and sufficient conditions that
\[\gamma(\omega) = \sum_{k=1}^\infty g_k(\omega)c_k \;\;\; (\omega>\omega_0)\]
should tend to a finite limit as $\omega\to\infty$ whenever
$\sum_{k=1}^\infty c_k = s$ is convergent are that
\begin{eqnarray}
&&(A) \sum_{k=1}^\infty \left|g_k(\omega) - g_{k+1}(\omega)\right|
\le M \;\;{\mbox{\rm\em for
every}}\;\; \omega>\omega_0,\nonumber\\
&&(B) \lim_{\omega\to\infty} g_k(\omega) = \beta_k \;\;{\mbox{\rm\em
for every fixed}}\;\; k. \nonumber
\end{eqnarray}

Moreover
\[(C) \lim_{\omega\to\infty} \gamma(\omega) = \beta_1 s +
\sum_{k=1}^\infty \left(\beta_k - \beta_{k+1}\right)(s_k -s),\]
where $s_k = \sum_{r=1}^k c_r$, and the existence of either side of
(C) implies that of the other, provided that $\sum_{k=1}^\infty c_k
$ converges to $s$.
\end{thm}

To exploit the theorem, we make the following identifications
\begin{eqnarray}
c_k &\rightsquigarrow& e^*_n f_n,\\
g_k(\omega) &\rightsquigarrow& \left(D(n)\right)^2,\;\;{\mbox{\rm
independent of $\omega$,}}
\end{eqnarray}
whereupon, $\gamma(\omega)$ is identified as the infinite sum
in~(\ref{condB}). We can now see that the simultaneous convergence
of the lhs of~(\ref{condA}) and~(\ref{condB}) are {\em not}
compatible because the condition $(A)$ of the theorem,
\begin{eqnarray}
(A) \rightsquigarrow \sum_{n=0}^\infty \left|\left(D(n)\right)^2 -
\left(D(n+1)\right)^2\right| \le M,
\end{eqnarray}
cannot be satisfied for any (Diophantine) polynomial $D(n)$, except
for a constant polynomial or for a finite truncation of the Hilbert
space, $\left(D(n)\right)^2=0$ for $n$ greater than some $N_0$, in
which case the infinite sum in $(A)$ turns into a finite sum and the
condition $(A)$ is then automatically satisfied with some finite and
sufficiently large $M$.	 Such a truncation is behind the
counterexample by Smith~\cite{Smith2005, kieuNew}.

The only possibility now left for the simultaneous convergence of
both~(\ref{condA}) and~(\ref{condB}) to zero is
\begin{eqnarray}
e^*_n f_n =0,\;\; {\mbox{\rm for all $n$ greater than some
$M_0$.}}\label{zero}
\end{eqnarray}
We next show that this last condition, in turn, contradicts the
normalisation~(\ref{norm}) of the eigenstates.

That $|f(t_0)\rangle$ is an eigenstate of ${\cal H}(t_0)$
in~(\ref{eigen1}) can be expressed explicitly as follows,
\begin{eqnarray}
\left(1-\frac{t_0}{T}\right)H_I |f(t_0)\rangle = \left(E_f(t_0)
-\frac{t_0}{T}H_P\right)|f(t_0)\rangle,\nonumber\\
\left(1-\frac{t_0}{T}\right)\left(a^\dagger a - \alpha a^\dagger -
\alpha^* a + |\alpha|^2\right) |f(t_0)\rangle = \left(E_f(t_0)
-\frac{t_0}{T}H_P\right)|f(t_0)\rangle, \label{21}
\end{eqnarray}
where we have substituted the explicit expression~(\ref{H_I}) for
$H_I$ with $K=1$.	 The rhs of~(\ref{21}) is {\em diagonal} in the
Fock basis; on the other hand, the $k$-th component of the lhs puts
some restriction on the values of $f_{k-1}$ and $f_{k+1}$ through
the action of $a^\dagger$ and $a$~\footnote{With $a^\dagger a
|k\rangle = k |k\rangle $, $a^\dagger |k\rangle = \sqrt{k+1}
|k+1\rangle $, and $a |k\rangle  = \sqrt k |k-1\rangle $.},
resulting in,
\begin{eqnarray}
-\left(1-\frac{t_0}{T}\right)\left(\alpha\sqrt k \;f_{k-1} +
\alpha^*\sqrt{k+1}\;f_{k+1}\right) = \left(E_f(t_0)
-\frac{t_0}{T}\left(D(k)\right)^2 -
\left(1-\frac{t_0}{T}\right)\left(k+|\alpha|^2\right)\right)f_k,
\nonumber\\
%\Rightarrow \;\; \alpha\sqrt k \;f_{k-1} +
%\alpha^*\sqrt{k+1}\;f_{k+1} &=& 0, \;\;{\rm if}\;\; f_k = 0 \;\;
%{\rm and}\;\; t_0<T.
\label{key}
\end{eqnarray}
for $k=0,1,\dots$	 It is worth stressing here that this
constraint~(\ref{key}) is only applicable to infinite dimensions.
With some simple modifications of the coefficients of $f_k$'s on the
lhs, it can also be applicable to a finitely truncated number of
dimensions, truncated to some $N_{\rm max}$, with appropriate
boundary conditions, such as the (anti)-periodic conditions,
$a^\dagger |N_{\rm max}\rangle = \mp c|0\rangle$ and $a |0\rangle =
\mp c^*|N_{\rm max}\rangle$ (with $c$ is some {\em non-zero} complex
number). In particular, the constraint~(\ref{key}) is not available
to the aruptly truncated condition, $a^\dagger |N_{\rm max}\rangle =
0 = a |0\rangle$, in the counterexample offered by
Smith~\cite{Smith2005}.

Now focusing on $f_n$, say, the condition~(\ref{zero}) implies that:
\begin{enumerate}
\item Either $f_n =0$ for $n$ greater than some $N_0>M_0$ (so that with no restriction on
$e_n$ for $n\ge N_0$ we still have $e^*_n f_n =0$):
\begin{quotation}
\noindent Applying~(\ref{key}) with $k=N_0$, we immediately have
$f_{N_0 - 1}=0$, for $t_0<T$.	 Then recursively applying~(\ref{key})
again but with $k=N_0-1$ and so on, we must have for $t_0<T$ and
{\em all} $n$,
\begin{eqnarray}
f_n = 0, \nonumber
\end{eqnarray}
contradicting the {\em non-zero} normalisation of $|f(t_0)\rangle$
in~(\ref{norm})!
\end{quotation}
\item Or more than two consecutive elements of $f_n$ vanishing, for instance
$(\ldots, f_{q-1}, 0, 0, f_{q+2}, \ldots)$ (and $e_{q-1}=0=e_{q+2}$,
etc.):
\begin{quotation}
\noindent Likewise, applying~(\ref{key}) with $k=q$ to have $f_{q -
1}=0$, for $t_0<T$, and with $k=q+1$ to have $f_{q+2}=0$.  Then
inductively applying~(\ref{key}) for larger and smaller values of
$k$, we must have for $t_0<T$ and {\em all} $n$,
\begin{eqnarray}
f_n = 0, \nonumber
\end{eqnarray}
again contradicting the {\em non-zero} normalisation of
$|e(t_0)\rangle$ in~(\ref{norm})!
\end{quotation}
\item Or $f_n$ asymptotically vanishes for every other value of $n$, for example
$(\ldots, f_{q-1}, 0, f_{q+1}, 0, f_{q+3},\ldots)$ (of course then
$e_n \to (\ldots, 0, e_{q}, 0, e_{q+2}, 0, e_{q+4},\ldots)$):
\begin{quotation}
\noindent Applying~(\ref{key}) with $k=q+1$, we have $f_{q+1}=0$ and
we are then back to the case above of more than two consecutive
elements of $f_n$ vanishing, which is once again in contradiction to
the {\em non-zero} normalisation of $|f(t_0)\rangle$
in~(\ref{norm}).

Alternative, we apply~(\ref{key}) with $k=q$ to obtain for
asymptotically large $k$,
\begin{eqnarray}
{k}\left|f_{k-1}\right|^2 =
(k+1)\left|f_{k+1}\right|^2\;\;\Rightarrow \;\;|f_{n}|^2 \sim
k\left|f_{k-1}\right|^2 O\left(\frac{1}{n}\right), \;\; {\rm
for}\;\; n>k\gg 1,\nonumber
\end{eqnarray}
contradicting the {\em convergence} of the infinite sum
$\sum_{n=k+1}^\infty |f_n|^2$~\footnote{As $\sum_{n=k+1}^\infty
|f_n|^2 \sim k\left|f_{k-1}\right|^2 \sum_{n=k+1} (1/n)$, and
$\sum_{n=k+1} (1/n)$ does not converge!} in the normalisation
condition~(\ref{norm}).
\end{quotation}

\item Or $f_n$ asymptotically vanishes
for only one component, $(\ldots, f_{q-2}, f_{q-1}, 0, f_{q+1},
f_{q+2},\ldots)$, then because of the constraint~(\ref{zero}), we
must have $(\ldots, 0, 0, e_q, 0, 0,\ldots)$ and are back to one of
the three cases above for $e_n$ with its contradiction to the {\em
non-zero} normalisation of $|e(t_0)\rangle$.
\item Likewise, for the case $f_n$ has no vanishing component
asymptotically, we can switch our attention to $e_n$, which must
asymptotically have some zero components because of~(\ref{zero}), to
derive some contradiction as in the above.
\end{enumerate}

All in all, the condition~(\ref{condition}) can never be violated.

Similar to those derived above for $K=1$, the contradictions for
higher numbers of variables $K$ should also follow essentially from
some key geometrical patterns of higher dimensional lattices, of
which each site is labelled by $K$ integer-valued coordinates.
Likewise, the contradictions can be manifest in general as the
violations of the non-zero normalisation condition, or of the
convergence of the series in the nomralisation, or both.

\section*{Summary} In~\cite{kieuFull}, we have argued that the
sufficient conditions in two dimensions for the measurement
probability for any $T$ is always less than one-half,
\begin{eqnarray}
|\langle \psi(T) | n_e \rangle|^2 &<& 0.5,\nonumber
\end{eqnarray}
where $|n_e\rangle$ is an eigenstate (Fock state) of, but not the
ground state of, the final Hamiltonian $H_P$, are that:
\begin{itemize}
\item $|\psi(0)\rangle = |\alpha\rangle$, the coherent ground state of the
initial Hamiltonian $H_I$;
\item $|\langle \psi(0) | n_e \rangle|^2 < 0.5$;
\item $\langle e(t)| H_P - H_I |f(t)\rangle \not=0$, for any pair of orthonormal instantaneous
eigenstates of ${\cal H}(t)$, and for all $0<t<T$.
\end{itemize}

Moving on to an infinite number of dimensions, we can, by
effectively reducing the problem to two dimensions through the
inductive consideration of pairs, each in turn, of orthonormal
instantaneous eigenstates, show that the conditions above are
sufficient even in the case of maximally constructive interference
among the various pairwise transitional amplitudes to any given
state.

On the one hand, we have shown above that the last sufficient
condition~(\ref{condition}) cannot be violated in a dimensionally
infinite Hilbert space or a finitely truncated space with
appropriate boundary conditions, ensuring that the measurement
probability at $T$ for {\em any} excited state, which is not
occupied at the initial time, cannot be more than one-half, for {\em
any} $T$.  On the other hand, we know from the quantum adiabatic
theorem that provided the time interpolation in~(\ref{calH}) is
sufficiently slow, that is, $T$ sufficiently large, the measurement
probability of the final ground state can be made arbitrarily close
to one, provided we start out in the initial ground state. Combining
these two, we thus arrive at an identification criterion for the
ground state of $H_P$ by a probability measure: {\em The final Fock
state that has a measurement probability of greater than one-half,
which can always be obtainable at some sufficiently large and finite
$T$, {\em is} the final ground state, assuming no degeneracy.}

This key result, nonconstructively proven, together with the quantum
adiabatic theorem are the main reasons behind the quantum
computability of some classical and recursive non-computable,
namely, Hilbert's tenth problem~\cite{kieuFull}.

\begin{acknowledgments}
I am indebted to Warren Smith for numerous email exchanges and for
his critical observations that led to a counterexample, which in
turn has led to the investigation above. I also wish to thank Peter
Hannaford, Toby Ord and Andres Sicard for support and discussion.
This work has been supported by the Swinburne University Strategic
Initiatives.
\end{acknowledgments}

\bibliography{c:/1data_16Apr05/papers/adiabatic}
\bibliographystyle{unsrt}

\end{document}